 \documentclass[a4paper,fleqn,usenatbib,onecolumn]{mnras}
\usepackage[T1]{fontenc}
\usepackage{ae,aecompl}
\usepackage{graphicx}   % Including figure files
\usepackage{amsmath}    % Advanced maths commands
\usepackage{amssymb}    % Extra maths symbols
\title[Electromagnetic radiation from  pair plasma inflow] 
 {Electromagnetic power induced from pair plasma
 falling into a rotating black hole II:
 An extensive WKB analysis in the slow rotation case
 }
 \author[Y. Kojima]{Yasufumi Kojima
 \thanks{%
 E-mail: ykojima-phys@hiroshima-u.ac.jp}\\ 
 Department of Physics, Hiroshima University, Higashi-Hiroshima, Hiroshima 
 739-8526, Japan}
\begin{document}
 \label{firstpage}
 \pagerange{\pageref{firstpage}--\pageref{lastpage}}
 \maketitle
\begin{abstract}
 We examine Poynting flux generation due to pair plasma accreting onto
a slowly rotating black hole. In particular, we consider the possibility of an outgoing flux at the horizon.
Our approach is based on a two-fluid model representing a collisionless 
pair plasma.
In the background, the plasma inflow is neutral and radial along 
the magnetic field lines of a split monopole in a Schwarzschild spacetime.
A combined mechanism of dragging by the black hole's spin 
and the Lorentz force produces charge separation and current flow,
and hence electric and toroidal magnetic fields.
By WKB analysis, two classes of solutions 
of perturbation equations for small black hole spin are identified: 
one is related to inward flux, and the amplitude inwardly increases;
the other generates an outward flux with a peak position around $r=3M$, 
tending to zero at the horizon.
The power induced by the two-fluid effect is inversely proportional to 
plasma density, and is small in almost all astrophysical situations. 
A magnetic vacuum region is located elsewhere for effective Poynting flux generation. 
\end{abstract}
%
%(1)%%%%%%%%%%%%%%%%%%%%%%%%%%%%%%%%%%%%%%%%%%%
\section{Introduction}
%%%%%%%%%%%%%%%%%%%%%%%%%%%%%%%%%%%%%%%%%%%%%%%
%
\citet{1977MNRAS.179..433B} (BZ) showed that there is an outgoing energy flux 
from a Kerr black hole.
For the last four decades, the BZ model has been discussed as a 
promising mechanism to power relativistic jets 
in active galactic nuclei, micro quasars and gamma ray bursts. 
The mechanism can be roughly summarized as follows. 
A spinning black hole distorts the poloidal magnetic field ${\vec B}_{p}$
and induces a poloidal electric field ${\vec E}_{p}$
and a toroidal magnetic field $ {\vec B}_{\phi}$, which generate an outward 
Poynting flux  ${\vec E}_{p} \times  {\vec B}_{\phi}/(4\pi)$
along the magnetic field lines threading the spinning black hole. 
Thus, the rotation energy of the spinning black hole is
electromagnetically extracted. 
The story is simple, but ambiguous in a sense.
Total energy flux $P$ through a radius in steady and axisymmetric 
electromagnetic field is expressed by integrating over a two-dimensional 
sphere: 
$ 4\pi P= -\int \Phi ,_{\theta} S d\theta d\phi$,
where $\Phi$ is an electric potential and $S$ is a poloidal current function.
The latter also describes $B_\phi$, with some geometric functions
\citep[See, e.g.,][] 
{1986bhmp.book.....T,2015MNRAS.454.3902K}.
This integral does not depend on the radius, 
when interaction between electromagnetic field and matter is 
completely neglected, like in force-free approximation.
It merely represents conservation of electromagnetic energy flow.
The mathematical expression represents the generation mechanism, 
but it should be note that not a dragging term 
$ {\vec \beta}_{\phi} \times {\vec B}_{p}$, but a gradient term  
${\vec \nabla \Phi}$ plays a crucial role, although stationary electric field 
generally consists of both terms.
It is therefore necessary to understand how the potential $\Phi$ is 
determined as the origin of electromagnetic power.
%

%2
  To be more specific, we consider the problem in the ideal MHD 
approximation, which is good in many astrophysical cases.
The electric potential $\Phi$ is constant along a 
poloidal magnetic line, which is characterized by a function $G$.
Its derivative $\Omega_{\rm F} \equiv d\Phi/dG$, which represents the angular 
velocity of the field line, is also constant along the line. 
In a steady problem, outward flux, say, 
at a certain radius far from a central object, 
leads to a horizon condition. 
Outgoing flux at the black hole horizon is mathematically 
shown to be possible only when 
$\Omega_{\rm F}$ is in a certain range, which
depends on the black hole spin.
Consider a trivial example:
when $\Omega_{\rm F} =0$ is fixed in the surroundings,
the flux is zero because $\Omega_{\rm F} =0$ everywhere.
Where and how is $\Omega_{\rm F} $ determined?
The issue is crucial for BZ process
  \citep[See, e.g.,][]{2014MNRAS.442.2855T,2016PTEP.2016f3E01T}.
On the analogy of a pulsar, the conversion of 
rotational energy to outward electromagnetic flux is sometimes discussed.
Magnetic lines are anchored on the central neutron star,
so that $\Omega_{\rm F} $ is fixed at the stellar surface.
The analogy is however inapplicable to the black hole magnetosphere,
since the black hole horizon is passive boundary.
That is, the horizon condition is physically 
determined as a result of exterior behavior.
There is then no good reason for any choice of $\Omega_{\rm F} $ 
at the horizon\citep[][]{1990ApJ...350..518P}.

Global steady-state force-free magnetospheres are 
modeled by numerically solving the relativistic 
Grad-Shafranov equation, in which there
are singular surfaces and careful treatment is necessary.
In this approach, $\Omega_{\rm F} $ is mathematically specified
to obtain a global solution
\citep[e.g.,][]{2004ApJ...603..652U,2005ApJ...620..889U,
2008PhRvD..78b4004T,2013ApJ...765..113C,2014ApJ...788..186N,
2015PhRvD..91f4067P,2016ApJ...816...77P,2017ApJ...836..193P,
2017PhRvD..95f3008T}.
For example, the force-free solution is obtained by expansion with 
respect to the black hole spin.
Split-monopole configuration is an exact solution in Schwarzschild spacetime, 
and $\Omega_{\rm F}$ is zero.
The first-order correction in $\Omega_{\rm F}$ is uniquely determined 
to avoid the divergence which appears at the horizon in solving a 
second-order magnetic function.
Similarly, 
\citet{2015PhRvD..91f4067P,2016ApJ...816...77P}
successfully calculated higher order corrections
using the horizon regularity and convergence constrain
in each order perturbation equation.
The mathematical treatment is correct, but 
there might be no consensus as to whether or not the divergence is 
seriously taken.
The force-free approximation breaks down near the horizon, since
the mass inertia of plasma becomes important in that region.
If so, the principle for determining $\Omega_{\rm F}$ 
is questionable in astrophysical meanings. 
The extension to the MHD case is necessary.
The formalism for stationary structure has already been given 
\citep[e.g.,][]{1990ApJ...363..206T,
1991PhRvD..44.2295N,1993PhyU...36..529B},
but it is quite difficult to obtain explicit solutions.
The problem of $\Omega_{\rm F}$ still remains,
because of no successful works within the formalism.
A time-dependent approach may be preferable to obtain the solutions.
Actually, general relativistic magnetohydrodynamic 
simulations provide very interesting models
\citep[e.g.,][]{2002Sci...295.1688K,2003ApJ...584..937V,
2004MNRAS.350.1431K,2005MNRAS.359..801K,
2006MNRAS.368.1561M,2009MNRAS.397.1153K,
2012MNRAS.423.3083M,2013MNRAS.436.3741P}.
Recently, very complicated but more realistic configurations
with very intense magnetic fields have been successfully analyzed.
The accretion inflow of matter, 
in disk flows called `` magnetically arrested disks'' (MADs)
and jet launching in the vicinity 
of a central black hole have been explored simultaneously.
However, it is hard to understand the origin of $\Omega_{\rm F} $
from these remarkable numerical results.

As far as we consider the problem of $\Omega_{\rm F}$ 
in a framework of MHD, the origin is likely to attribute to the boundaries.
Otherwise, a model to fix $\Omega_{\rm F}$ should be designed elsewhere
in the interior.
Here, we take a different approach to the origin of
$\Omega_{\rm F}$ (or $\Phi$).
We investigate whether or not there is an intrinsic mechanism in 
Kerr spacetime to produce it.
If so, is the value at the horizon
an appropriate one for the outward flux?
For this purpose, we have to study plasma flows consistent 
with the electromagnetic fields in a two-fluid model
  \citep[e.g.,][]{2009MNRAS.398..271K,2014MNRAS.438..704B,
2015MNRAS.446.2243P,2017JCAP...05..041P},
where $\Phi $ is no longer a constant along a magnetic field line.  
A previous paper\citep[][hereafter referred to Paper I]{2015MNRAS.454.3902K} 
provided a general framework for an
axially symmetric and stationary system around a Kerr black hole. 
It is difficult to consistently solve the whole set of equations, which is 
a coupled system of four partially differential equations.
To obtain a definite solution, the effects of first-order slow spin 
were considered. Namely, in a Schwarzschild spacetime,
the flow is radial along a split-monopole magnetic field.
There is no charge density or current flow, and hence
the electric and toroidal magnetic fields vanish everywhere. 
By taking into account the slow rotation, 
perturbations of these fields are induced. However, the analysis in Paper I
is incomplete, since there is an error in eq. (42) of section 3.3.
Here we correct that error and further study 
the electric-field generation problem in the slow rotation regime.

  This paper is organized as follows.
We first summarize our basic equations, which contain a large dimensionless parameter. 
These equations are not easy to solve numerically, so we approximate them by leading-order terms
and discuss this limitation in section 2.
Results based on WKB analysis are given in section 3.
Finally, section 4 is our conclusion.
We use units in which $c=G_{\rm N}=1$.

%(2)%%%%%%%%%%%%%%%%%%%%%%%%%%%%%%%%%%%%%%%%%%%
\section{Model and Formulation}
%%%%%%%%%%%%%%%%%%%%%%%%%%%%%%%%%%%%%%%%%%%%%%%
%2.1%%%%%%%%%%%%%%%%%%%%%%%%%%%%%%%%%%%%%%%%%%%
\subsection{Basic equations}
%%%%%%%%%%%%%%%%%%%%%%%%%%%%%%%%%%%%%%%%%%%%%%%
A general formalism was given in Paper I for
axisymmetric stationary states of
two-component plasma flows consistent with
electromagnetic fields in a Kerr spacetime.
Perturbation equations 
with respect to the slow rotation of the black hole were also derived
in order to evaluate the effect of the black hole spin.
Here we also limit ourselves to the slow rotation regime
and summarize the relevant equations below.

Schwarzschild spacetime
with the first order rotational correction is given by
\begin{equation}
ds^2=-\alpha^2dt^2+\alpha^{-2}dr^2
       +r^2d\theta^2 +r^2\sin^2\theta d\phi^2
        -2\omega r^2\sin^2\theta dt d\phi,
\end{equation}
where
\begin{equation}
\alpha^2=1-\frac{2M}{r},
~~ \omega=\frac{2M^2a_*}{r^3}.
\end{equation}
Here, $M$ is a mass, and 
$a_*$ is dimensionless small spin parameter.

Magnetic fields are assumed to be a  
split monopole with typical field strength $B_0$
and its perturbations,
which are generally described by two functions,
$\delta G(r,\theta)$ and $\delta S(r,\theta)$:
\begin{equation}
[B_{\hat{r}},B_{\hat{\theta}}, B_{\hat{\phi}}]
=\left[\frac{B_{0}M^2}{r^2}
+ \frac{\delta G,_\theta}{r^2 \sin\theta} 
,~
-\frac{\alpha \delta G,_r}{r \sin\theta}, 
 ~
 \frac{\delta S}{\alpha r\sin \theta}\right] .
\label{Bcomp}
\end{equation}
Electric fields are the first-order quantity 
described by a function $\delta\Phi (r,\theta)$: 
\begin{equation}
[E_{\hat{r}},E_{\hat{\theta}}, E_{\hat{\phi}}]
=\left [
  -\delta\Phi,_r ,~ 
-\frac{1}{\alpha r}(\delta\Phi,_\theta - \omega B_{0} M^2 \sin\theta),
~0  \right] ,
\label{Ecomp}
\end{equation}
where the second term in $E_{\hat{\theta}}$ represents dragging a radial
magnetic field in the azimuthal direction.

The plasma is modeled as a cold collisionless fluid of
particles with mass $m$ and electric charge $\pm e$.
The flow of each component is described by the stream function 
$F_{\pm} =F_0 +\delta F _{\pm}$,
where the background flow described by $F_0 $ 
is radial along the magnetic monopole field.
There is no net charge density or current flow in the background, 
to be consistent with the electromagnetic fields.
The radial flow velocity is 
$ v _{\hat{r}} = -(2/x)^{1/2}$, irrespective of species, and the common number density is 
$ n_{0}= \lambda n_{c}(2x^2(x-2))^{-1/2} $,
where $x=r/M$, $n_{c} = B_0/(4\pi e M)$
and $\lambda$ is a dimensionless number.
The perturbation  $\delta F _{\pm}$ is separated into two modes, 
`even' $\delta F _{+} =\delta F _{-} $
and `odd' $\delta F _{+} =-\delta F _{-} $.
In the latter, a poloidal current 
and a toroidal magnetic field are induced
since the current is produced by the difference between the two streams:  
$\delta S = 4 \pi e (\delta F _{+} -\delta F _{-} ) \ne 0$.
The flow directions in the meridian plane are opposite
 $\delta v^{+} _{\rm p} =-\delta v^{-}_{\rm p} $, 
whereas those in the azimuthal direction are the same 
 $\delta v^{+} _{\phi} =\delta v^{-}_{\phi} $.
The Lorentz forces for each component are opposite in the $\theta$ direction:
$ \pm e ( \delta v^{\pm }_{\hat\phi}  B_{\hat r}) $.
The number densities do not balance, 
$\delta n_{+ } =-\delta n_{- } $, so that charge density 
$\delta \rho_{e} = e (\delta n _{+} -\delta n _{-} ) \ne 0$ is induced.
The non-vanishing toroidal current is the second order
$\delta j_{\phi} $
$=e( \delta n_{+ }\delta v^{+} _{\phi} 
- \delta n_{-}\delta v^{-} _{\phi} ) $ in this mode.
We restrict ourselves to odd mode perturbations 
$\delta F _{+}$ $=-\delta F _{-}$, and neglect
the perturbation of the magnetic function $\delta G$ in eq. (\ref{Bcomp}). 
Furthermore, the angular part is decoupled by the following forms: 
\begin{equation}
\delta \Phi =  B_0 M h(r) \cos \theta, ~
 \delta F _{\pm}
= \pm (\lambda n_c M^2/2) p(r) \sin^2 \theta , ~
\delta S =\lambda  B_0 M p(r) \sin^2 \theta  ,
\label{eqnPFS}
\end{equation}
since the slow rotation corresponds to a
dipole perturbation with spherical harmonic index $l=1$.
With these approximations, a system of four partial differential 
equations 
(Poisson's equation, the Biot-Savart equation,
and an equation for each stream function) 
is reduced to two pairs of ordinary differential 
equations for $h$ and $p$
%%%
\footnote{ This set of equations is obtained from eqs. (41) and (42)
in paper I, but there was a mistake in eq. (42).
The errata for paper I are given in the Appendix.}:
%%%
\begin{equation}
\left[ \kappa^{-2} \frac{d^2}{dx ^2}
+ U_0 - \kappa^{-2}U_2 \right] \left( x h \right) 
 -\frac{\sqrt{2}k^{1/2}}{\kappa}\alpha^{-2} x^{-5/4}
\left(x^{3/4} p \right) = 
\frac{4 a_{*}}{ \kappa^2\alpha^{2}x^{4}},
\label{perteqnhh}
\end{equation}

\begin{equation}
\left[
 \kappa^{-2} \frac{d^2}{dx ^2} -V_0 +  \kappa^{-2}V_2
\right]\left(x^{3/4} p \right)
 -\frac{1}{k^{1/2}\kappa}\alpha^{-2} x^{-5/4}\left( x h \right) 
=
\frac{2 a_{*}}{ k^{1/2}\kappa\alpha^{2}x^{13/4}},
\label{perteqnpp}
\end{equation}
where $x=r/M$, and the potential terms are divided into
\begin{equation}
U_0 =(1/\sqrt{2x}) , 
U_2 =2/(\alpha^2x^{2}), 
%~~
%\label{eqnWKBUPT}
\end{equation}
\begin{equation}
V_0 =\sqrt{2}x^{-3/2} \alpha^{-2} v,
~~~
v \equiv 1- k^{-1}(2x)^{-3/2} \alpha^{2} ,
~~
V_2 =3/(16x^{2}).
%%
%\label{eqnWKBPVT}
\end{equation}
In eqs. (\ref{perteqnhh}) and (\ref{perteqnpp}),
two parameters are involved\footnote{
There are three parameters in a two-fluid model
\citep[e.g.][]{2014MNRAS.438..704B}, but
one associated with relaxation time
vanishes due to our collisionless approximation.
}.
One is a dimensionless plasma frequency 
$\kappa ^2 \equiv \omega_{p} ^2 M^2$
$= 4\pi e^2 (\lambda  n_{c})M^2/m$, where the typical number density 
is estimated with multiplicity $\lambda $
and `Goldreich-Julian density' $ n_{c} \equiv B_{0}/(4\pi e M)$.
In astrophysical situations, $\kappa $ is very large,
$\kappa \sim 10^{10}$.
Another parameter $k$ represents the ratio of the rest mass energy 
density of pairs to the electromagnetic energy density:
$k =  (m \lambda n_{c})/(B_{0} ^2 /4 \pi)$.
When $k \gg 1$, hydrodynamical effects, such as pressure,
are important and energy flow by matter dominates.
In this case, our treatment is no longer valid.
However, our concern is magnetically dominated flow, so 
we do not consider the large $k$ case.
The multiplicity $\lambda$ is expressed as
$\lambda = k^{1/2} \kappa$.
The reasonable condition $\lambda >1$ leads to $\kappa^{-2} < k$.
We also consider a lower bound of $k$, and
take $k_{c} \approx 2.3 \times 10^{-2}$ 
in order to simplify our argument.
The number $ k_{c} $ will be derived in the next subsection, 
and hence the range of $k$ is of order $10^{-2}$-$10^{0}$,
which covers the astrophysically interesting cases.
%

%%2.2%%%%%%%%%%%%%%%%%%%%%%%%%%%%%%%%%%%%%%%%%%%%%
\subsection{Further approximation and limitation}
%%%%%%%%%%%%%%%%%%%%%%%%%%%%%%%%%%%%%%%%%%%%%%%%%%

%
It is a natural approximation to neglect higher order terms, except the derivative terms,
with $\kappa^{-n}$ $(n\ge 2)$ in eqs. (\ref{perteqnhh}) and (\ref{perteqnpp}), 
 because $ \kappa \gg 1$. 
The equations are then reduced to decoupled equations for
$h$ and $p$, and the solution can be easily obtained.
Explicit forms will be given in the next section.
Setting $\kappa^{-2}U_{2}=0$ in eq. (\ref{perteqnhh}),
the solution $h$ is oscillatory, since
the potential $U_0$ is positive definite.   
Similarly, eq. (\ref{perteqnpp}) without 
the term $\kappa^{-2}V_2$ gives 
an exponential type solution for $p$,
as long as  $V_0$ is positive.   
The condition for this is given by $k > k_c\approx 2.3\times 10^{-2} $.
When  $k < k_c$, the potential $V_0$ becomes negative in a 
range $r_1< r < r_2$, and the function becomes oscillatory there.
The whole solution is obtained
by matching functions at $r_1 $ and $r_2 $.
We expect that such a solution is possible 
for only a particular value of $k$, namely, an eigenvalue,
and requires more careful treatment.
Our discussion is mainly limited to the range $k > k_c$.

Ignoring formally small terms proportional to $\kappa^{-2}$
is a great simplification, but 
restricts the applicable range at the same time.
The term $\kappa^{-2}U_{2}$ in eq. (\ref{perteqnhh})
increases toward the horizon because
$U_{2} \propto \alpha^{-2}$.
The potential term $U_{0} -\kappa^{-2}U_{2}$ 
becomes negative inside $r_c$, where a turning radius
$r_{c}$ is approximated as
$1-(2M/r_{c})=\alpha^2 (r_c) \approx \kappa ^{-2} \ll 1$.
The resulting solution changes from oscillatory to exponential 
growth/decay behavior across $r_c$.
The term $\kappa^{-2}V_2$ in eq. (\ref{perteqnpp})
has a minor effect since
it is small everywhere.
Thus, we may safely ignore this term.
The approximation to set $\kappa^{-2}U_{2}=\kappa^{-2}V_{2}=0$ 
is limited to the range of $r > r_{c} (\approx 2M)$.
This limitation also affects the inner boundary condition.
A regularity condition for eqs. (\ref{perteqnhh}) and (\ref{perteqnpp}) 
at the horizon is given by
%.
\begin{equation}
k^{1/2} p  +\frac{h}{\kappa } +\frac{ a_{*}}{ 4 \kappa} =0 .
\label{ZnajekCNhpa}
\end{equation}
Divergent terms in the limit of $\alpha \to 0$ 
are canceled in each equation,  
when eq. (\ref{ZnajekCNhpa}) is satisfied.
This is nothing but
the incoming wave condition of the electromagnetic fields,
or the Znajek condition near the horizon
$ B_{\hat{\phi}}=-E_{\hat{\theta}} $
\citep{1978MNRAS.185..833Z,1986bhmp.book.....T}.
Equation (\ref{ZnajekCNhpa}) is derived
in terms of the first-order perturbed functions $h$ and $p$,
from eqs. (\ref{Bcomp}) and (\ref{Ecomp}).
The regularity condition is not the same in the approximated system 
with $\kappa^{-2}U_{2}=0$. 
Thus, the condition (\ref{ZnajekCNhpa}) is not necessary
at the inner boundary $r \to 2M$,
although $r_{c}$ is numerically close to the horizon $2M$.
The condition (\ref{ZnajekCNhpa})
is a passive one near the horizon, and is automatically satisfied
in a regular system for $2M \le r \le r_{c} $. 
%

%(3)%%%%%%%%%%%%%%%%%%%%%%%%%%%%%%%%%%%%%%%%%%%%
\section{WKB Analysis}
%%%%%%%%%%%%%%%%%%%%%%%%%%%%%%%%%%%%%%%%%%%%%%%%
%3.1%%%%%%%%%%%%%%%%%%%%%%%%%%%%%%%%%%%%%%%%%%%%
\subsection{Solutions for homogeneous equations}
%%%%%%%%%%%%%%%%%%%%%%%%%%%%%%%%%%%%%%%%%%%%%%%%
We solve the homogeneous equations 
with the approximations $\kappa^{-2}U_2=\kappa^{-2}V_2=0$
in eqs. (\ref{perteqnhh}) and (\ref{perteqnpp}).
The solutions also describe the 
perturbations in a Schwarzschild black hole.
We seek an approximate WKB solution of the form
$ p \propto \exp( \kappa W(x) )$ and $ h \propto \exp( \kappa W(x) )$,
where $\kappa (\gg 1)$ is a large number.
Substituting these into eqs.~(\ref{perteqnhh})--(\ref{perteqnpp}), 
we find the leading-order solutions correct to order $ \kappa ^{-1}$.
The four independent solutions (two pairs) given below are denoted 
by $h^{\pm} _{n}$ and $p^{\pm} _{n}$.

%%1%%
A pair of type I solutions is given in terms of $x=r/M$ by
\begin{eqnarray}
h^{\pm} _{{\rm I}}  &= x^{-1} U_0 ^{-1/4}
 \exp\left(\pm i \kappa \int^{r/M} U_0 ^{1/2} d{\bar x} \right)
\nonumber
\\
&=  2^{1/8} x^{-7/8} \exp(\pm i \kappa_1  x^{3/4} ),
\label{eqn.typuu1}
\end{eqnarray}
\begin{equation}
p^{\pm} _{{\rm I}} = - 2^{1/2} k^{-1/2}\kappa^{-1} x^{-1/2}Q^{-1}
h^{\pm} _{{\rm I}}, 
\label{eqn.typuu}
\end{equation}
where
\begin{equation}
Q = 1 -2^{-1/2} k^{-1} \alpha ^2 x^{-5/2}
\end{equation}
and $\kappa_1 = (2^{7/4}/3)\kappa $. The overall constant 
from the integral is adjusted to normalize the solution in eq.~(\ref{eqn.typuu1}).
This solution represents $p^{\pm} _{{\rm I}} $
$\ll h^{\pm} _{{\rm I}}$ in the large $\kappa$ limit:
the larger function $h^{\pm} _{{\rm I}}$ is solely
determined by eq. (\ref{perteqnhh}), and constrains
the smaller $p^{\pm} _{{\rm I}}$ through eq. (\ref{perteqnpp}).
The typical oscillatory scale is $\sim \kappa^{-1} M \sim \omega_{p}^{-1}$, and 
changes with the spatial distribution of the background number density.

%%2%%
Another pair of type II solutions is
\begin{eqnarray}
p^{\pm} _{{\rm II}} &=
x^{-3/4} V_0 ^{-1/4} \exp\left( \pm \kappa
 \int^{r/M} V_0 ^{1/2} d{\bar x} \right) ,
\label{eqn.typv0}
\\
&= 2^{-1/8} \alpha^{1/2} x^{-3/8} v^{-1/4}
\exp\left( \pm g(x, \xi) \right) ,
\end{eqnarray}
\begin{equation}
h^{\pm} _{{\rm II}}  = 2 k^{1/2} \kappa^{-1} x^{-1}  
Q^{-1} p^{\pm} _{{\rm II}} ,
\label{eqn.typvv}
\end{equation}
where
\begin{equation}
 g(x, \xi)=  2^{1/4}\kappa
\int_{\xi } ^{x} \alpha^{-1} {\bar x}^{-3/4} v^{1/2} d {\bar x} ,
\label{eqn.gexpII}
\end{equation}
and $\xi $ is a constant.
This solution represents $h^{\pm} _{{\rm II}} $
$\ll p^{\pm} _{{\rm II}}$ in the large $\kappa$ limit.
The larger $p^{\pm} _{{\rm II}}$ is solely
determined by eq. (\ref{perteqnpp}), and constrains
the smaller $h^{\pm} _{{\rm II}}$ through eq. (\ref{perteqnhh}).
The situation is opposite to that of the type I solution.
The two classes of solutions are clearly decoupled,
since the coupling terms in eqs. (\ref{perteqnhh}) and (\ref{perteqnpp})
decrease with $\kappa^{-1}$.
The electrostatic perturbation dominates in the type I solution, 
whereas the fluid perturbation dominates in the type II solution.

\begin{figure}
%\centering
%
\includegraphics[scale=0.65]{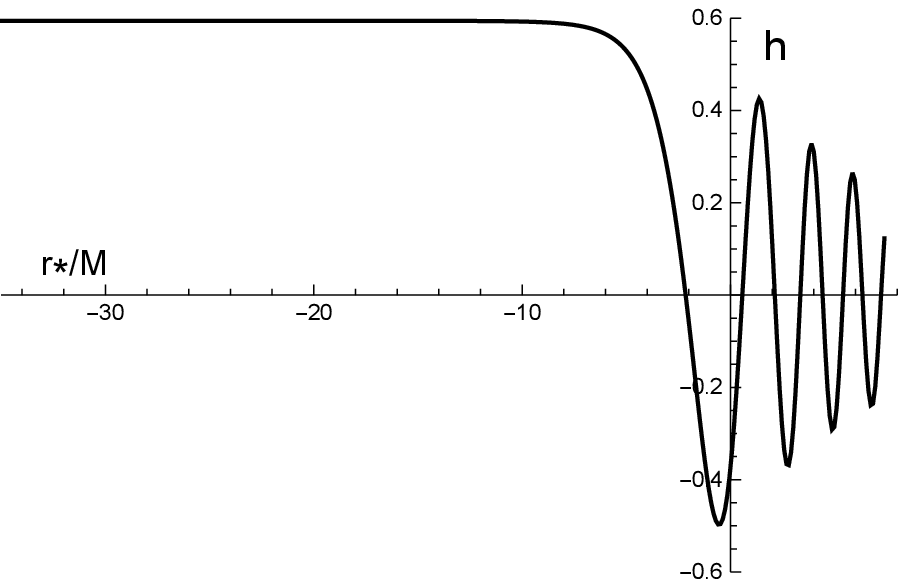}
\includegraphics[scale=0.65]{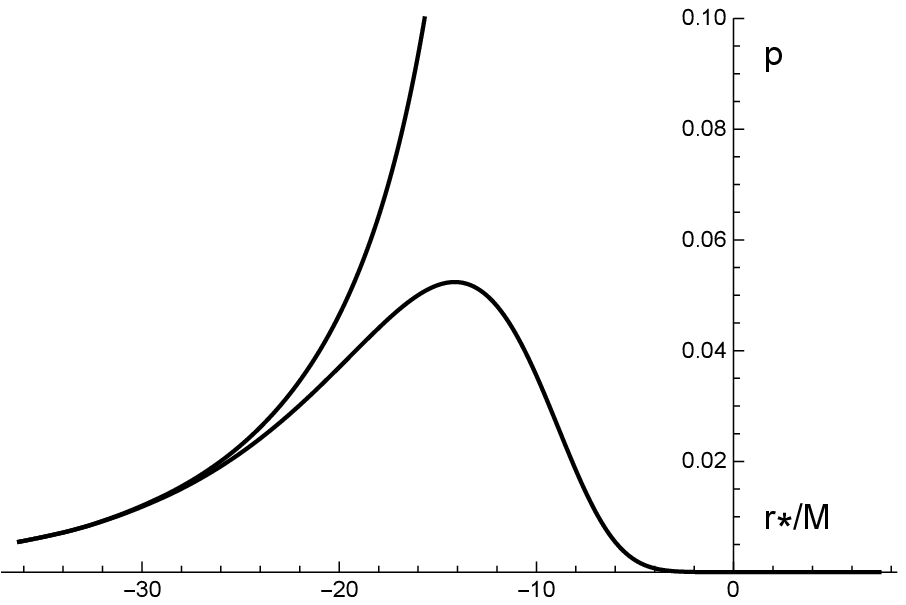}
\caption{
\label{Fig.1ab}
 WKB solutions as a function of $r_*$.
The type I solution $h$ is shown in the left panel.
Growing and decaying solutions $p$ in 
the type II case are shown in the right panel.
The parameters are chosen as $\kappa  =10$, $k=0.1$.
The valid range $r>r_{c}$ discussed in section 2.2
corresponds to $r_{*}/M > -7.2$ for $\kappa =10$.
}
\end{figure}

  Figure \ref{Fig.1ab} shows these functions with the
tortoise coordinate $r_{*}$$\equiv r+ 2M\ln(r/2M-1)$.
The two types are clearly distinguished by their 
functional behavior:
one is oscillatory, and the other grows or decays exponentially.
The function $h_{\rm I}$ in left panel of Fig.~\ref{Fig.1ab}
is oscillatory outside a certain radius.
The oscillation is determined by $\kappa_{1}x^{3/4}$, and 
the wavelength becomes small with increasing $\kappa (\propto \kappa_{1} )$.
The oscillation is very rapid for a realistic value of $\kappa$.
The envelope of the oscillations inwardly increase as $r^{-7/8}$, but
the oscillation stops and 
the function tends to a constant as $r_{*} \to -\infty$.
It is found that the transition point is very close to the 
critical radius $r_{c}$. 
Thus, asymptotic behavior for $r < r_{c}$ is meaningless, and the 
function there should be exponentially growing or decaying 
by the correction term $\kappa^{-2}U_{2}$ in eq. (\ref{perteqnhh}). 
The function $p_{\rm II}$ in the right panel of Fig.~\ref{Fig.1ab}
tends to zero at the horizon with a factor $\alpha^{1/2}$, 
whereas it is exponentially growing or decaying
for large $r_*$.

%%%%%%%%%%%%%%%%%%%%%%%%%%%%%%%%%%%%%%%%%%%%%%%
%%%%% FIG1  %%%%%
%%%%%%%%%%%%%%%%%%%%%%%%%%%%%%%%%%%%%%%%%%%%%%%

%%
As shown in eqs. (\ref{eqn.typuu}) and (\ref{eqn.typvv}),
the two functions $h$ and $p$ are connected by a function $Q^{-1}$
in both type I and type II solutions.
The function $Q$ becomes zero, when 
$k < k_{0}\approx 1.54\times 10^{-2}(<k_{c})$.
There is a divergence at $Q=0$, but
this is an artifact of neglecting $\kappa^{-2}$ terms.
In our consideration limited to $k >k_c $,
the function is approximated as $Q \approx 1$.
Thus, we have $h p <0 $ in the type I solutions, 
whereas $h p >0 $ in the type II solutions.
The relative sign is important for the direction 
of energy flow, as discussed in section 3.3.

%%3.2%%%%%%%%%%%%%%%%%%%%%%%%%%%%%%%%%%%%%%%%%%%%%%%%
\subsection{Solutions with spacetime dragging effect}
\begin{figure}
\centering
 \includegraphics[scale=0.80]{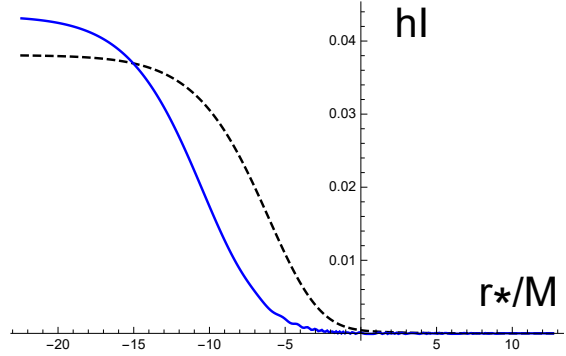}
\caption{
\label{fig2}
The special solution $h^{\rm S} _{{\rm I}} \kappa/a_*$ 
is displayed as a function of $r_*$.
The solid line represents the function for $\kappa=10^2$, whereas
the dashed one is for $\kappa=10^1$.
The growth point shifts inward
with the increase of $\kappa$.
The result hardly depends on the parameter $k$, 
so long as $k > k_{c}$. Here $k=0.1$ is used.
}
\end{figure}

%%%%%%%%%%%%%%%%%%%%%%%%%%%%%%%%%%%%%%%%%%%%%%%%%%%%% 
%%
  A general solution of eqs. (\ref{perteqnhh}) and (\ref{perteqnpp})
without the source terms is expressed by linear combinations 
of four functions as $ h = \sum c^{\pm} _{n} h^{\pm} _{n}(x)$, 
and $ p = \sum c^{\pm} _{n} p^{\pm} _{n}(x)$.
The solution of the inhomogeneous equation
is obtained by varying the coefficients $ c^{\pm} _{n} $ as
 $ h = \sum c^{\pm} _{n} (x) h^{\pm} _{n}(x)$, 
and $ p = \sum c^{\pm} _{n} (x) p^{\pm} _{n}(x)$.
Putting these forms into eqs. (\ref{perteqnhh}) and (\ref{perteqnpp}),
we have
%%%%
\begin{equation}
\frac{1}{\kappa}  
\frac{d c^{\pm} _{{\rm I}} }{dx}  =
\mp \frac{i}{2} \frac{J_{{\rm I}}}{U_0^{1/4}} 
\exp\left[ \mp i\kappa \int ^{r/M} U_0^{1/2} d x ^\prime \right] ,
\label{c1dfeq}
\end{equation}
\begin{equation}
\frac{1}{\kappa}  
\frac{d c^{\pm} _{{\rm II}} }{dx}  =
\pm \frac{1}{2} \frac{J_{{\rm II}}}{V_0^{1/4}} 
\exp\left[ \mp \kappa \int ^{r/M} V_0^{1/2} d x ^\prime \right] ,
\label{c2dfeq}
\end{equation}
where
\begin{equation}
J_{{\rm I}} =
\frac{4 a_{*}}{ \kappa^2\alpha^{2}x^{4}}
\left(1- x^{-1/4}Q^{-1}\right),
\label{c1soc}
\end{equation}
\begin{equation}
J_{{\rm II}} =\frac{2 a_{*}}{ k^{1/2}\kappa\alpha^{2}x^{13/4}} .
\label{c2soc}
\end{equation}
%%%%
Here we considered the leading order terms with respect to $\kappa^{-n}$.
We integrate eq. (\ref{c1dfeq}) with boundary condition
$c^{\pm} _{{\rm I}}=0$ at large radius. A particular solution of the 
inhomogeneous equation is given by
\begin{equation}
 h^{\rm S}_{{\rm I}} = 2^{9/4} a_* \kappa^{-1} x^{-7/8}  
 \int _{x} ^{\rm out} 
\left( 1-\xi^{-1/4}Q^{-1} \right) \alpha^{-2} \xi^{-31/8}
\sin\left[ \kappa_1 (\xi^{4/3} -x^{4/3} ) \right] d\xi
\label{eqn.wkb.inhs}
\end{equation}
and $p^{\rm S}_{{\rm I}}$ can be obtained by the relation (\ref{eqn.typuu}).
%
%% Fig2 %%
%
Figure \ref{fig2}  shows the function $h^{\rm S} _{{\rm I}} \kappa/a_*$ 
in eq. (\ref{eqn.wkb.inhs})
for $\kappa=10^{1},10^{2}$ with $k=0.1$.
As $x=r/M$ decreases, $h^{\rm S} _{{\rm I}} $
grows from zero and approaches a constant.
The calculation is carried out for $\kappa$ not too large, 
since the cost of the calculation increases with $\kappa$. However, a general property can be inferred: as $\kappa$ increases, the growing point shifts to a smaller radius
and the asymptotic constant of $h^{\rm S} _{{\rm I}} $
is almost proportional to $\kappa^{-1}$.
The highly oscillatory region contributes little to
the integral (\ref{eqn.wkb.inhs}) due to cancellation.
The growth is thus related to the termination of the oscillation
or a ``frozen star'' property near the black hole horizon.
The growing point of $h^{\rm S} _{{\rm I}} $ and
the critical radius $r_{c}$ are close to each other,
and both move inward with increasing $\kappa$.
The saturation region in Fig. \ref{fig2} may be out of range, 
although the integral (\ref{eqn.wkb.inhs}) is carried 
for small $r$ to demonstrate the functional behavior.
The solution $h^{\rm S} _{{\rm I}} $ depends on 
the parameter $k$ through $Q$, but this dependence is weak since $Q \approx 1$.
Thus, we estimate $h^{S} _{{\rm I}} \propto \kappa^{-1} a_{*}$
and $p^{S} _{{\rm I}} \propto k^{-1/2}\kappa^{-2}a_{*} $.
These correspond to  $E_{p} \propto \kappa^{-1} a_{*}$ and
$B_{\phi} \propto \kappa^{-1} a_{*}$
in the electromagnetic perturbations.

%%%%%%%%%%%%%%%%%%%%%%%%%%%%%%%%%%%%%%%%%%%%%%%
%%%%% FIG2 %%%%%
%%%%%%%%%%%%%%%%%%%%%%%%%%%%%%%%%%%%%%%%%%%%%%%

\begin{figure}
\centering
 \includegraphics[scale=0.80]{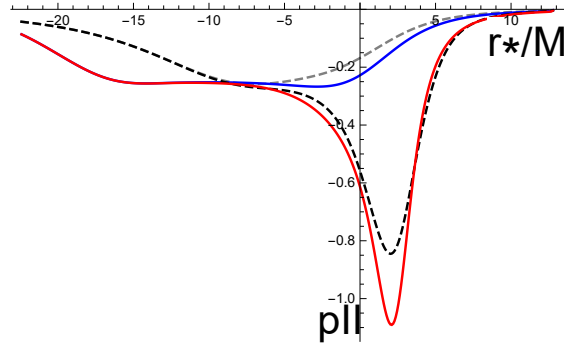}
\caption{
\label{fig3}
The special solution $p^{\rm S} _{{\rm II}} k^{1/2}\kappa/a_*$ 
is displayed as a function of $r_*$.
The solid lines represent the functions for $\kappa=10^2$, whereas
dashed ones are for $\kappa=10^1$.
The two lower curves with the minimum $\sim -1$ 
are for $k=2.5 \times 10^{-2}$, whereas the upper ones are for $k=0.1$.
}
\end{figure}

%%2%%
  We integrate eq.~(\ref{c2dfeq}) with two boundary conditions:
$c^{+} _{{\rm II}}=0$ at a large radius, and
$c^{-} _{{\rm II}}=0$ from an inner point.
%By integrating eq.~(\ref{c2dfeq}), 
A particular solution of the 
inhomogeneous equation can be written in a concise form as
%\begin{eqnarray}
\begin{equation}
%&&
 p^{\rm S}_{{\rm II}} = - \frac{a_* \alpha^{1/2} }
{2^{1/4}k^{1/2} x^{3/8} v^{1/4}}  
 \int _{\rm in} ^{\rm out} 
\alpha^{-3/2} \xi^{-23/8} v^{-1/4} 
\exp\left[ -|g(x, \xi) | \right] d \xi ,
\label{eqn.wkbII.inps}
\end{equation}
where $g(x, \xi) $ is defined in eq. (\ref{eqn.gexpII}).
This is a method to solve inhomogeneous equations in terms of 
a Green function constructed by the WKB approximation.
\citep[see, e.g.,][]{1999amms.book.....B}.
%
%% Fig3 %%
%
The integral in eq. (\ref{eqn.wkbII.inps})
provides a value of order $\kappa^{-1}$, so 
the normalized function $p^{\rm S} _{{\rm II}} \kappa k^{1/2}/a_*$
is shown for $k=0.1, 2.5\times 10^{-2}$ 
and $\kappa=10^{1},10^{2}$ in Fig.\ref{fig3}.
The function shows a steep minimum around $r_*/M \approx 2.5$, 
($r/M \approx 3.3$) for $k=2.5\times 10^{-2}$.
It becomes deeper as $k \to k_{c} $.
This sharp minimum comes from the function $v$, 
which has a minimum at $x=r/M=10/3$.
The solutions are damped toward the horizon
by a factor $\alpha^{1/2}$ in the homogeneous solutions
$p^{\pm} _{{\rm II}}  $.
Overall the functions scale 
as $p^{S} _{{\rm II}}  \propto k^{-1/2}\kappa^{-1}a_{*}$,
and $h^{S} _{{\rm II}}  \propto \kappa^{-2} a_{*} $
through eq. (\ref{eqn.typvv}), except for a region around
the accidental point $r/M =10/3$.
These behaviors correspond to $E_{p} \propto \kappa^{-2} a_{*}$ and
$B_{\phi} \propto \kappa^{0} a_{*}$
in the electromagnetic perturbations.
In the limit of the ideal MHD case ($\kappa^{-1} \to 0$),
a toroidal magnetic field is generated by dragging, 
whereas the electric potential remains zero
as imposed in the outer boundary condition.

%%%%%%%%%%%%%%%%%%%%%%%%%%%%%%%%%%%%%%%%%%%%%%%
%%%%% FIG3 %%%%%
%%%%%%%%%%%%%%%%%%%%%%%%%%%%%%%%%%%%%%%%%%%%%%%

%%3.3%%%%%%%%%%%%%%%%%%%%%%%%%%%%%%%%%%%%%%%%%%%%%%%%
\subsection{Electromagnetic energy flow}
%%%%%%%%%%%%%%%%%%%%%%%%%%%%%%%%%%%%%%%%%%%%%%%%%%%%%

We now discuss the Poynting power induced by black hole spin.
The electromagnetic energy flow originates from the 
product of the induced electric and toroidal magnetic fields.
The energy through a sphere of radius $r$ is calculated as
(paper I):
\begin{equation}
P_{{\rm em}}(r)=- \int (\sqrt{-g} T_{{\rm em} ~t} ^r ) d\theta d\phi 
=-\frac{1}{2}\int _{r={\rm const.}} 
(\delta \Phi,_{\theta} \delta S ) d\theta .
%
%\label{EMpower}
\end{equation}
Using the first-order perturbations, $h$ and $p$ in eq. (\ref{eqnPFS}), we have 
\begin{equation}
P_{{\rm em}}= \frac{2}{3} k^{1/2} \kappa h p  (B_0 M)^2 . 
%\label{EMpower}
\end{equation}
The sign of $h p$ determines the direction of the energy flow.
When $k > k_c$ $h^{\rm S} _{{\rm I}} $ and $p^{\rm S} _{{\rm I}} $ 
give $P_{{\rm em}}<0$, that is, inflow toward the black hole.
There is a lower limit $r_c$, which is introduced 
due to our approximation of neglecting the 
higher order terms proportional to $\kappa^{-2}$. 
The interior solution in $2M <r< r_c$ 
is of the exponential type due to the negative potential 
($U_{0} -\kappa^{-2}U_{2} <0 $ in eq. (\ref{perteqnhh})),
and can be obtained by matching 
interior and exterior solutions across $r_{c}$.
We do not explicitly work out the full solution, but the sign of both functions
 $h^{\rm S} _{{\rm I}} $ and $p^{\rm S} _{{\rm I}} $ 
is likely to keep it in the exponential form inside $r_{c}$.
Thus, the energy is still inflowing at the horizon.
When $k < k_{0}$, the function $Q$ becomes negative
between $r_{1}$ and $r_{2}$,
where the two radii are approximately given by
$1-(2M/r_{1}) \approx 2^3 k \ll 1$ and 
$r_{2}/M \approx  (2k^2)^{-1/5} \gg 1$.
In this limited region, outgoing power is induced
because $h^{\rm S} _{{\rm I}} p^{\rm S} _{{\rm I}} >0 $. 
Both $r_c$ and $r_1$ are close to $2M$, but 
we have $r_c < r_1 $ 
since $\kappa^{-2} \ll 1$ and $ k \approx {\mathcal O}(1)$ 
in astrophysical situations. 
Therefore, the flux becomes negative again between $r_c $ and $ r_1 $,
and the situation is the same near the horizon.
Thus, the solutions $h^{\rm S} _{{\rm I}} $ and $p^{\rm S} _{{\rm I}} $ 
do not result in an outgoing flux at the horizon.

\begin{figure}
\centering
 \includegraphics[scale=0.80]{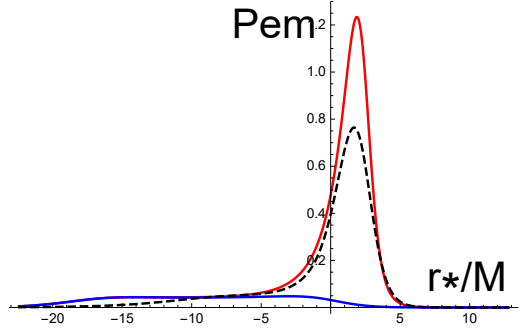}
\caption{
\label{fig4}
Outgoing Poynting flux $P_{{\rm em}}\kappa^{2}/(a_{*}B_{0}M)^{2}$
as a function of $r_*$.
The top solid line shows the result for 
the solution $p^{\rm S} _{{\rm II}} $ and  $h^{\rm S} _{{\rm II}} $
with $k=2.5\times 10^{-2},\kappa=10^2$;
the dashed line is for $k=2.5\times 10^{-2},\kappa=10^1$; and
the lower flat line is for $k=0.1,\kappa=10^2$.
}
\end{figure}

%%2%%
We next discuss a set of $h^{\rm S} _{{\rm II}} $ and $p^{\rm S} _{{\rm II}} $,
which satisfies $ h p > 0$  when  $k  > k_c$.
In this case an outgoing flux is generated.
The function $P_{\rm em}$ is shown in Fig. \ref{fig4}.
A sharp peak is located around $r_{*}/M \approx 2.5~(r/M \approx 3$),
and $P_{\rm em}$ goes to zero on both sides
($r_{*}/M \to \pm \infty$). 
A Poynting flux is generated inside the peak.
Namely, 
material energy is converted to electromagnetic energy in that region.
Outside the peak, the conversion is in the opposite direction.
The decrease of $P_{\rm em}$ at large $r$ depends on
the outer condition, $p^{\rm S} _{{\rm II}} \to 0$.
The decrease of $P_{\rm em}$ toward the horizon
comes from the functional behavior,
$ p _{{\rm II}} \propto \alpha^{1/2} $, 
$ h _{{\rm II}} \propto \alpha^{1/2} $ as $\alpha  \to 0 $.
The Poynting power is in order of magnitude  
$P_{{\rm em}} \approx  \kappa^{-2} \times P_{{\rm BZ}}$,
where $P_{{\rm BZ}} (\approx  (a_{*} B_{0} M)^2)$
is the Blandford and Znajek power. The magnitude of $P_{{\rm em}}$ 
is small due to a small factor $\kappa^{-2}$.
The power can also be written as $P_{{\rm em}} \propto n_{0}^{-1} B_{0} ^2$,
which decreases with number density $n_{0}$.
%

%%%%%%%%%%%%%%%%%%%%%%%%%%%%%%%%%%%%%%%%%%%%%%%
%%%%% FIG4 %%%%%
%%%%%%%%%%%%%%%%%%%%%%%%%%%%%%%%%%%%%%%%%%%%%%%

%%3.4%%%%%%%%%%%%%%%%%%%%%%%%%%%%%%%%%%%%%%%%%%%%%%%%
\subsection{Growth and decay of charge separation}
%%%%%%%%%%%%%%%%%%%%%%%%%%%%%%%%%%%%%%%%%%%%%%%%%%%%%

In the previous subsection, the power is shown to be  
generated not around $ r \approx 2M$
but around $ r \approx 3M$.
The generation mechanism fails toward the horizon.
We study how and where a black hole spin affects 
neutral radial flow in background.
Lorentz force, in particular, its $\theta$-component is very important
to produce non-radial spatial deviation:
\begin{equation}
\delta f^{\pm} _{\hat \theta}= \pm e
( \delta E_{\hat \theta} +\delta v^{\pm} _{\hat \phi} B_{\hat r} -
   v_{\hat r} \delta B_{\hat \phi} )
=
\pm \frac{m \kappa^2}{M k}
\left[ 
\frac{k^{1/2}}{\kappa \alpha x}
\left( h+ \frac{2a_{*}}{ x^{3}}\right)
-\frac{\alpha}{2 x^{3}} p
+\frac{2^{1/2}k }{\alpha x^{3/2}} p \right] \sin \theta,
\label{LforceTh}
\end{equation}
where three terms in the first expression are
explicitly written down by $p$ and $h$ in the second one
\footnote{
Equations (\ref{Bcomp}),(\ref{Ecomp}),
$ \delta v^{\pm} _{\hat \phi} = -k^{-1/2}\kappa p \sin\theta/(2\gamma_0 x)$
and $\alpha=\gamma_{0}^{-1}$ from eq.(38) in Paper I are used.
}. 
We at first consider the behavior of $\delta f^{\pm} _{\hat \theta}$
in far region, where $r\gg 2M $ and  $\alpha \approx 1$.
The second term $\delta v^{\pm} _{\hat \phi} B_{\hat r}$
is dominant for type II solution ($h \ll p $),
in a reasonable parameter range $\kappa^{-2} < k < 1$.
Black hole drags the plasma in azimuthal direction   
irrespective of their electric charge ($\pm e$), and the Lorentz force acts
in an opposite direction with respect to the fluid species.
This mechanism causes spatial unbalance between two fluid components,
and leads to nonzero charge density and current flows.
Nonzero electric potential
$\delta \Phi$ and toroidal magnetic field $\delta B_{\hat \phi}$
are thus produced.
Toward black hole horizon ($\alpha \to 0$), 
third term increases due to a factor $\alpha^{-1} $.
This term has an opposite sign compared with the second one.
Resultant toroidal magnetic field $\delta B_{\hat \phi}$
suppress growth of the $\theta$-motion.
As $r$ further approaches the horizon, the force (\ref{LforceTh}) 
seems to diverge. A relation between $h$ and $p$ 
in the coefficient of  $\alpha^{-1} $ is nothing but 
the Znajek condition, eq.(\ref{ZnajekCNhpa}).
So they should be canceled, and $\delta f^{\pm} _{\hat \theta}$ vanishes.
The flow becomes radial near the horizon.
The outward electromagnetic power also decreases there. 
%  

%(4)%%%%%%%%%%%%%%%%%%%%%%%%%%%%%%%%%%%%%%%%%%%
\section{Conclusion}
%%%%%%%%%%%%%%%%%%%%%%%%%%%%%%%%%%%%%%%%%%%%%%%

A poloidal electric field, in particular 
its potential part is essential to Poynting flux 
in a stationary and axially symmetric system.
Once the potential is set to zero, for example, by a certain mechanism
at an outer radius, the potential and resultant Poynting flux
are both zero everywhere in a region threaded by magnetic field lines, 
since the potential is constant in the ideal MHD approximation. 
We have attempted to explore the origin 
of finite flux by a two-fluid effect, where
the potential is no longer constant along the magnetic field.

There is a large dimensionless number contained in the formalism. 
That is, a ratio between microscopic scale of plasma and
macroscopic scale of a black hole. 
This fact hinders the numerical integration for a realistic value.
Using WKB analysis, we could classify modes and
estimate the parameter dependence in a simple model.
One mode describes an energy inflow toward the horizon,
and the amplitude is finite there.
The other describes an outgoing energy flow, and the luminosity
has a sharp peak at some distance from the horizon.
The magnitude decreases inward to zero, and the mode
does not yield outgoing flow from the horizon.
  Furthermore, the resultant Poynting power is very small: it is reduced
by a small factor $\kappa^{-2}$, where
$\kappa$ is a dimensionless plasma frequency, compared with
the BZ power ($\approx  (a_{*} B_{0} M)^2$).
With increasing $\kappa$, that is, increasing plasma number density,
the ideal MHD condition becomes a better approximation, and 
the electric field vanishes 
to become consistent with the outer boundary value.
The luminosity ($ \propto n_{0}^{-1} B_{0} ^2$) 
decreases with an increase of plasma density $ n_{0}$.
In this study, we found outward Poynting flux 
induced by the black hole spin, but failed to
apply it in astrophysical situation.
The two-fluid effect was not so important.
However, this conclusion may be related to 
the simple model considered here.
It is necessary to consider the effect on more elaborate models.  
The two-fluid effect is effective in a low-density region,
so that a successful model requires such 
a magnetic vacuum region elsewhere in the black hole magnetosphere.
The region is also related to 
a pair creation region or an origin of wind
\citep[e.g.,][]{2000NCimB.115..795B,2008ASSL..355.....P,
2010PhyU...53.1199B,2012PASJ...64...50O,2015PASJ...67...89O}.
In their models, the position is proposed by some arguments.
Another drawback in present model is the first-order limit
of a Kerr black hole spin.
The ergo-radius coincides with the horizon, so that
there is no region inherent in the black hole spin.
A rapidly rotating black hole significantly affects plasma
flows and may produce
an extremely low-density region, where
the two-fluid effect is efficient.
Further study is challenging.

%%Appendix%%%%%%%%%%%%%%%%%%%%%%%%%%%%%%%%%%%%%%%%
\section*{Appendix:errata in paper I}
%%%%%%%%%%%%%%%%%%%%%%%%%%%%%%%%%%%%%%%%%%%%%%%%%%
There is a mistake in eq. (42) of paper I\citep{2015MNRAS.454.3902K}, 
which leads 
to an incorrect functional behavior for type II solutions.
The coefficient $\alpha^{-2}s^{3/2}$
in front of $dp/ds$ in eq. (42) should be changed to $s^{3/2}$.
The factor $\alpha^{-2}$ leads to 
$p^{\pm } _{{\rm II}} \propto \alpha$ and
$h^{\pm } _{{\rm II}} \propto \alpha^{-1}$
in paper I, but the correct behaviors
are $p^{\pm } _{{\rm II}} \propto \alpha^{1/2}$ and
$h^{\pm } _{{\rm II}} \propto \alpha^{1/2}$.
See eqs. (\ref{eqn.typv0})--(\ref{eqn.typvv}) in this paper.
The special solution $p^{{\rm S}} _{{\rm II}}$ (eq. (53) of paper I)
is also wrong, and is corrected as eq. (\ref{eqn.wkbII.inps})
in this paper.
All figures with $p^{{\rm S}} _{{\rm II}}$ or 
$h^{{\rm S}} _{{\rm II}}$  in paper I are wrong.
%%%%%%%%%%%%%%%%%%%%%%%%%%%%%%%%%%%%%%%%%%%%%%%%%%%%%%%%%%%%

%%%%%%%%%%%%%%%%%%%%%%%%%%%%%%%%%%%%%%%%%%%%%%%%%%
\section*{Acknowledgements}
%%%%%%%%%%%%%%%%%%%%%%%%%%%%%%%%%%%%%%%%%%%%%%%%%%
This work was supported  by JSPS KAKENHI Grant Number JP26400276.

%%%%%%%%%%%%%%%%%%%% REFERENCES %%%%%%%%%%%%%%%%%%
   \bibliographystyle{mnras}
   \bibliography{kojima17Nov}
%%%%%%%%%%%%%%%%%%%%%%%%%%%%%%%%%%%%%%%%%%%%%%%%%%%%%%%%
 \end{document}